\documentclass[12pt,a4paper,reqno]{article}
\usepackage{amsmath}\usepackage{amssymb} 
\usepackage{amsthm}

\renewcommand{\(}{\left(} 
\renewcommand{\)}{\right)}

\renewcommand{\S}{\mathcal{S}}

\newcommand{\ox}{\otimes}
 
\newcommand{\<}{\langle}
\renewcommand{\>}{\rangle} 
\newcommand{\half}{\tfrac{1}{2}}

\theoremstyle{plain} 
\newtheorem{theorem}{Theorem}

\newtheorem{corollary}{Corollary}
 
\numberwithin{equation}{section}
\begin{document} 
\title{\bfseries{Multipartite generalisation of the Schmidt
decomposition}} 
\author{\\\\ H. A. Carteret$^1$, A. Higuchi$^2$ and A. Sudbery$^3$\\ 
\small \emph{Dept. of Mathematics, University of York, 
Heslington, York, YO10 5DD, U.K.} \\
\small email: 
$^1$hac100@york.ac.uk, $^2$ah28@york.ac.uk, $^3$as2@york.ac.uk} 

\date{27 June 2000, revised 15 November 2000}

\maketitle


\begin{abstract}
We find a canonical form for pure states of a general multipartite
system, in which the constraints on the coordinates (with respect to a
factorisable orthonormal basis) are simply that certain ones vanish and
certain others are real. For identical particles they are invariant under
permutations of the particles. As an application, we find the 
dimension of the generic local equivalence class.
\end{abstract}
 
\section{Introduction}

Recently considerable attention \cite{Acin, BrunCohen, HilaryNSme,
Hilaryme, Grasslinv, Kempeinv, NoahSandu, Makhlin, localinv, Thapliyal,
Woot:quant, Woot:tang} has been devoted to the problem of
describing the equivalence classes of states of a composite quantum
system, where two states are regarded as equivalent if they are related
by a unitary transformation which factorises into separate
transformations on the component parts (a \emph{local} unitary
transformation). One approach to this problem is to specify a canonical
form for states under local unitary transformations. For pure states of
two-part systems, such a canonical form is given by the Schmidt
decomposition 
\begin{equation}
|\Psi\> = \sum_i\alpha_i|\phi_i\>|\psi_i\>
\end{equation}
where the $|\phi_i\>$ are a set of orthogonal states of the first
subsystem, the $|\psi_i\>$ are a set of orthogonal states of the second
subsystem, and the $\alpha_i$ are positive real numbers. The state
$|\Psi\>$ is thus expanded in terms of a factorisable basis of two-part
states so that the number of non-zero coefficients is minimal. Ac\'{\i}n et
al \cite{Acin} have shown that there is an expansion of states of three
qubits which has a similar property. In this
note we will demonstrate such a decomposition for pure states of an $n$-part
system, where the dimensions of the individual state spaces are finite
but otherwise arbitrary. 
Then we use this to find the dimension of the generic local equivalence
class.
We will also comment on the
relation of this decomposition to other proposed canonical forms.

\section{The generalised Schmidt decomposition}

We first state and prove the generalisation of the Schmidt decomposition for a
multipartite state in which all the individual state spaces have the
same dimension, since this is considerably simpler than the general case:

\begin{theorem} Let $|\Psi\>$ be a state vector in an $n$-fold tensor
product space $\S_1 \ox \cdots \ox \S_n$ where 
$\dim\S_1 = \cdots =\dim\S_n = d \geq 2$ and $n \geq 3$. Then for
$r=1,\ldots,n$ there is a basis $\{|\psi^{(r)}_i\>: i=1,\ldots,d\}$ of
$\S_r$ such that in the expansion
\begin{equation}
|\Psi\>=\sum_{i_1\cdots i_n}c_{i_1\cdots i_n}|\psi^{(1)}_{i_1}\>\cdots
|\psi^{(n)}_{i_n}\>
\end{equation} 
the coefficients $c_{i_1\cdots i_n}$ have the following properties:
\end{theorem}
\begin{enumerate}
\item $c_{jii\cdots i} = c_{iji\cdots i} = \cdots = c_{ii\cdots
ij} = 0$ if $1\leq i <j \leq d$;
\item $c_{i_1\cdots i_n}$ is real and non-negative if at most one of the
$i_r$ differs from $d$;
\item $|c_{ii...i}| \geq |c_{j_1...j_n}|$ if $i \leq j_r$, $r=1,...,n.$
\end{enumerate}

\begin{proof}
Consider the real-valued function $\left|\langle \Psi|
\left(|\phi^{(1)}\rangle \cdots |\phi^{(n)}\rangle \right)\right|^2$
defined for unit vectors $|\phi^{(r)}\rangle$ lying in the unit sphere
$S^{2d-1}$ in $\mathcal{H}_r.$  As $\left(|\phi^{(1)}\rangle,
\ldots,|\phi^{(n)}\rangle \right)$ varies over the compact space
$S^{2d-1} \times \cdots \times S^{2d-1},$ this function attains a
maximum at some point $\left(|\psi_1^{(1)}\rangle, \ldots,
|\psi_1^{(n)}\rangle \right).$  Let $\left\{|\psi_i^{(r)}\rangle:
i=1,\ldots,d \right\}$ be any orthonormal basis of $\mathcal{H}_r$
containing $|\psi_1^{(r)}\rangle,$ and expand $|\Psi\rangle$ as in the
statement of the theorem.  Since $\left|\langle \Psi
|\left(|\psi_1^{(1)}\rangle \cdots |\phi^{(r)}\rangle \cdots
|\psi_1^{(n)}\rangle \right)\right|^2$ is stationary at
$|\phi^{(r)}\rangle = |\psi_1^{(r)}\rangle$ for variations of
$|\phi^{(r)}\rangle$ on the unit sphere,
\begin{equation}
 c_{1 \ldots 1 j 1 \ldots 1} = \langle \Psi | \left( 
   |\psi_1^{(1)} \rangle \cdots |\psi_1^{(r-1)}\rangle
   |\psi_j^{(r)}\rangle |\psi_1^{(r+1)} \rangle
    \cdots |\psi_1^{(n)}\rangle \right) = 0 \quad \text{for} \quad j > 1.
\end{equation}
Next, find the maximum of $\left|\langle \Psi|\left( |\phi^{(1)} \rangle
\cdots |\phi^{(n)}\rangle \right) \right|^2$ as $|\phi^{(1)} \rangle,
\ldots, |\phi^{(n)} \rangle$ vary over unit vectors orthogonal to
$|\psi_1^{(1)} \rangle, \ldots, |\psi_1^{(n)} \rangle$ respectively.  
Suppose the maximum occurs at $\left( |\psi_2^{(1)}\rangle, \ldots,
|\psi_2^{(n)} \rangle \right).$  Then, as before, in any expansion of
$|\Psi\rangle$ in terms of orthonormal bases of the $\mathcal{H}_r$
containing $|\psi_1^{(r)} \rangle$ and $|\psi_2^{(r)} \rangle, \quad
(r=1,\ldots,n),$ the coefficients will satisfy 
\begin{equation}
 c_{2\ldots 2j2 \ldots 2}=0 \quad \text{for} \quad j>2.
\end{equation}
Continuing in this way, we define the basis vectors $|\psi_i^{(1)}
\rangle, \ldots, |\psi_i^{(n)} \rangle$ for $i=1, \ldots, d-1.$  Then
the last basis vector $|\psi_{d}^{(1)} \rangle$ of $\mathcal{H}_1$ is
determined up to phase.

The reality conditions can be imposed as follows.   Choose the phase of
the basis element $|\psi_i^{(r)} \rangle, \quad i=1,\ldots,d-1$ so that
\begin{equation}
 \arg (c_{d \ldots did \ldots d}) = 0
\end{equation}
where the index $i$ occurs in the $r$th place, and then fix 
$c_{d \ldots d}$ by choosing the phase of $|\psi_d^{(1)} \rangle.$
\end{proof}

The general form
of the theorem is rather more complicated than the above; to state it,
we need to define the following sets of $n$-tuples. 

Let $(I_1, \ldots, I_N)$ be the set of 
$(n-1)$-tuples $(i_1,\ldots , i_{n-1})$ with $1 \leq i_r \leq d_r$,
excluding those
of the form $(i,\ldots,i)$ with $1\leq i < d_1$ and those of the form
$(d_1,\ldots,d_r,i,\ldots,i)$ with $d_r\le i<d_{r+1}$ and 
$1\leq r \leq n-2$. We order these $(n-1)$-tuples in
lexicographical order, so that $I_N=(d_1,\ldots,d_{n-1})$. 
Let $D=d_1\cdots d_{n-1}$, so that $N=D-d_{n-1}+1$. We define
$A$ to be the set of $n$-tuples $(i_1,\ldots,i_n)$ with
$(i_1,\ldots,i_{n-1})= I_k$ and $i_n=d_{n-1}+l$ where
$1 \leq k \leq \text{min}(N, d_n - d_{n-1})$ and 
$k \leq l \leq d_n-d_{n-1}$.

We also define the following sets of $n$-tuples:
\begin{align*}
B_1 = &\{(d_1,\ldots, d_{r-1}, j, d_{r+1}, \ldots, d_{n-1}, d_{n-1}):\,
1 \leq j < d_{r}\,,\ \ 
1 \leq r \leq n-2\},\\
B_2 = &\{ (d_1, \ldots, d_{n-2}, j, d_{n-1}):\,
1 \leq j < d_{n-2}\}, \\
B_3 = &\{ (d_1, \ldots, d_{n-2}, j, 1) :\,
d_{n-2} \leq j \leq d_{n-1}\},\\
B_4 = &\{ (i,i,\ldots, i):\, 2 \leq i \leq d_1\}\\
&\cup\{(d_1, \ldots, d_{r}, i, \ldots, i):\,
1\leq r < n-1,\;d_{r} < i \leq d_{r+1}\}\\
&\cup\{(d_1, \ldots, d_{n-1},i):\, d_{n-1} < i \leq \text{min}(D,d_n)\}.
\end{align*}

Then $A, B_1,\ldots,B_4$ are all disjoint. In terms of these sets, the 
general Schmidt decomposition can be stated as follows:

\newpage
\begin{theorem} Let $|\Psi\>$ be a state vector in an $n$-fold tensor
product space $\S_1\ox\cdots\ox\S_n$ where dim$\S_r=d_r$, with $2\leq d_1\le
\cdots \leq d_n$, and $n \geq 3$. Then for
$r=1,\ldots,n$ there is a basis $\{|\psi^{(r)}_i\>: i=1,\ldots,d_r\}$ of
$\S_r$ such that in the expansion
\begin{equation}
|\Psi\>=\sum_{i_1\cdots i_n}c_{i_1\cdots i_n}|\psi^{(1)}_{i_1}\>\cdots
|\psi^{(n)}_{i_n}\>
\end{equation}
the coefficients $c_{i_1\cdots i_n}$ have the following properties:
\end{theorem}
\begin{enumerate}
\item $c_{ii\cdots ijii\cdots i} = 0$ if $1\leq i < d_1$ and $i<j$.
\item $c_{d_1\cdots d_r ii\cdots ijii\cdots i} = 0$ if $d_r\leq i<d_{r+1}$ and
$i<j$, $1\leq r\leq n-2$.
\item $c_I=0$ for every $n$-tuple index $I$ in the set $A$.
\item The coefficients with indices in the sets
$B_i$ $(i=1,\ldots, 4)$
are real and non-negative. 
\item For $i = 1,\ldots, d_{n -1}$, define
\[ 
R_i = |c_{d_1\cdots d_r i \cdots i}|
\] 
where $r$ is such that $d_r < i \leq d_{r+1}$. Then
\[ 
R_1 \geq \cdots \geq R_{d_{n-1}}.
\]
\end{enumerate}

\begin{proof}
Consider the real-valued function
$\left|\<\Psi|\(|\phi^{(1)}\>\cdots|\phi^{(n)}\>\)\right|^2$ defined for unit
vectors $|\phi^{(r)}\>$ lying in the unit sphere $S^{2d_r -1}$ in $\S_r$.
As $(|\phi^{(1)}\>,\ldots ,|\phi^{(n)}\>)$ varies
over the compact space $S^{2d_1-1}\times\cdots\times S^{2d_n-1}$, this
function attains a maximum at some point 
$(|\psi^{(1)}_1\>,\ldots ,|\psi^{(n)}_1\>)$.
Let $\{|\psi^{(r)}_i\>:\;i=1,\ldots ,d_r\}$ be any orthonormal basis of $\S_r$
containing $|\psi^{(r)}_1\>$, and expand $|\Psi\>$ as in the statement
of the theorem. Since $\left|\<\Psi|\(|\psi^{(1)}_1\>\cdots |\phi^{(r)}\>\cdots
|\psi^{(n)}_1\>\)\right|^2$ 
is stationary at $|\phi^{(r)}\>=|\psi^{(r)}_1\>$ for
variations of $|\phi^{(r)}\>$ on the unit sphere,
\begin{equation}
c_{1\cdots 1j1\cdots 1} =
\<\Psi|\(|\psi^{(1)}_1\>\cdots|\psi^{(r-1)}_1\>|\psi^{(r)}_j\>
|\psi^{(r+1)}_1\>\cdots|\psi^{(n)}_1\>\)= 0 \quad\text{ for }j>1.
\end{equation}

Next, find the maximum of $\left|\<\Psi|\(|\phi^{(1)}\>\cdots 
|\phi^{(n)}\>\)\right|^2$  as
$|\phi^{(1)}\>,\ldots,|\phi^{(n)}\>$ vary over unit vectors orthogonal to
$|\psi^{(1)}_1\>,\ldots ,|\psi^{(n)}_1\>$ respectively.
Suppose the maximum occurs at $\(|\psi^{(1)}_2\>,\ldots
,|\psi^{(n)}_2\>\)$. Then, as before, in any expansion of $|\Psi\>$ in
terms of orthonormal bases of the $\S_r$ containing
$|\psi^{(r)}_1\>$ and $|\psi^{(r)}_2\>$ $(r=1,\ldots ,n)$, 
the coefficients will satisfy
\begin{equation}
c_{2\cdots 2j2\cdots 2} = 0 \quad \text{for} \quad j>2.
\end{equation}
Continuing in this way, we define the basis vectors
$|\psi^{(1)}_i\>,\ldots ,|\psi^{(n)}_i\>$ for $i = 1, \ldots ,d_1-1$.
Then the last basis vector $|\psi^{(1)}_{d_1}\>$ of $\S_1$ is determined
up to phase. Next, if the dimensions $d_i$ are not all equal, we maximise
$\left|\<\Psi|\(|\psi^{(1)}_{d_1}\>|\phi^{(2)}\>\cdots|\phi^{(n)}\>\)\right|^2$
as $|\phi^{(2)}\>,\ldots,|\phi^{(n)}\>$ vary orthogonally to the basis
vectors already determined, to find basis vectors
$|\psi^{(2)}_{d_1}\>,\ldots,|\psi^{(n)}_{d_1}\>$;
then we find $|\psi^{(2)}_i\>, \ldots ,|\psi^{(n)}_i\>$ for
$i=d_1+1,\ldots,d_2-1$, and hence $|\psi^{(2)}_{d_2}\>$;
then $|\psi^{(3)}_i\>, \ldots ,|\psi^{(n)}_i\>$ for $i=d_2,\ldots,d_3-1$;
and so on until we have $|\psi^{(n-1)}_i\>,|\psi^{(n)}_i\>$ for
$i=d_{n-2},\ldots ,d_{n-1}-1$. The maximisation at each step implies that the
coefficients satisfy (2).
 
Now, to fix the last $d_n-d_{n-1}+1$ basis elements of $\S_n$, we choose 
a set $I_1=(i_1,\ldots,i_{n-1})$ 
of $n-1$ indices which is not of the form
$(i,\ldots,i)$ with $1 \leq i < d_1$ or 
$(d_1,d_2,\ldots,d_r,i,i,\ldots,i)$ with $d_r\leq i < d_{r+1}$
(so that $c_{i_1\cdots i_{n-1}j}$
has not yet been set to zero for any $j$), and maximise
$\left|\<\Psi|\(|\psi^{(1)}_{i_1}\>
\cdots|\psi^{(n-1)}_{i_{n-1}}\>|\phi^{(n)}\>\)\right|^2$
with respect to vectors $|\phi^{(n)}\>$ orthogonal to
$|\psi^{(n)}_1\>,\ldots,|\psi^{(n)}_{d_{n-1}-1}\>$, 
thus finding $|\psi^{(n)}_{d_{n-1}}\>$;
then, choosing a different
index set $I_2=(j_1,\ldots,j_{n-1})$, maximise
$\left|\<\Psi|\(|\psi^{(1)}_{j_1}\>
\cdots|\psi^{(n-1)}_{j_{n-1}}\>|\phi^{(n)}\>\)\right|^2$
with respect to vectors $|\phi^{(n)}\>$ orthogonal to
$|\psi^{(n)}_1\>,\ldots,|\psi^{(n)}_{d_{n-1}}\>$; and so on until we
have either exhausted the possible index sets $I_1,I_2,\ldots, I_N$ or run
out of space in which to vary the vector $|\phi^{(n)}\>$. The coefficients
will then satisfy (3).

The reality conditions (4) can be imposed as follows. (By \emph{using} a
basis vector $|\psi\>$ to \emph{fix} a coefficient $c$ we mean changing the
phase of $|\psi\>$ to make $c$ real and non-negative.) First use 
$|\psi^{(n)}_1\>$ to fix $c_{d_1 \cdots d_{n-1}\,1}$; then use
$|\psi^{(n)}_{d_{n-1}}\>$ to fix $c_{d_1 d_2\cdots d_{n-1} d_{n-1}}$;
then use $|\psi^{(r)}_j\>$ $(r=1,\ldots,n-2;\;j=1,\ldots,d_r-1)$ to fix 
the coefficients in the set $B_1$; then use $|\psi^{(n-1)}_j\>$
$(j=1,\dots,d_{n-1}-1)$ to fix the coefficients in $B_2$ and $B_3$; and
finally use $|\psi^{(n)}_j\>$ $(j=2,\ldots,\text{min}(d_n,D)$ excluding
$j=d_{n-1}$) to fix the remaining coefficients in $B_4$.
\end{proof}

This result can be expressed in terms of active transformations, with
respect to fixed orthonormal bases $\{|\theta^{(r)}_i\>\}$ of the state
spaces $\S_r$, as follows. Two states $|\Psi_1\>, |\Psi_2\>$ in
$\S_1\ox\cdots\ox\S_n$ are said to be \emph{locally equivalent} if
\[ |\Psi_1\> = (U_1\ox\cdots\ox U_n)|\Psi_2\>
\]
where $U_r$ is a unitary transformation acting on $\S_r$. Then we have 
\begin{theorem} For $n \geq 3$, 
any state $|\Psi\>\in \S_1\ox\cdots\ox\S_n$ is locally
equivalent to a state $\sum c_{i_1\ldots i_n}|\theta^{(1)}_{i_1}\>\cdots
|\theta^{(n)}_{i_n}\>$ where the coefficients $c_{i_1\ldots i_n}$ have the
properties 1--5 stated in Theorem 2.
\end{theorem}

As an example, we note the canonical form for a state of three
qubits,
\begin{equation}
a|000\> + b|011\> + c|101\> + d|110\> + e|111\>
\end{equation}
with $b,c,d,e$ real.  Ac\'{\i}n et al 
\cite{Acin} have investigated tripartite states using different canonical forms
which, like the canonical form proposed here, have five non-zero 
coefficients of which four are real.  

We note that for a state expressed in terms of a fixed orthonormal basis
$\{|\theta_i^{(r)}\>\}$ as
\[
|\Psi\> = \sum t_{i_1\cdots i_n}|\theta^{(1)}_{i_1}\>\cdots
|\theta^{(n)}_{i_n}\>,
\]
the coefficients of the basis elements $|\psi_1^{(r)}\> =
\sum u^{(r)}_{i_r}|\theta^{(r)}_{i_r}\>$ defined in the first step of the
above proof are the solutions of the generalised (nonlinear) ``singular 
value'' equations
\begin{equation}\label{E-L}
\sum_{i_1\cdots i_{r-1},i_{r+1}\cdots i_n}\overline{t_{i_1\cdots i_n}}
u^{(1)}_{i_1}\cdots
u^{(r-1)}_{i_{r-1}}u^{(r+1)}_{i_{r+1}}\cdots u^{(n)}_{i_n} = \lambda
\overline{u^{(r)}_{i_r}}
\end{equation}
where the Lagrange multiplier $\lambda$  is such that $|\lambda|^2$ is
the maximal value of $|\<\Psi|\(|\phi^{(1)}\>\cdots |\phi^{(n)}\>\)|^2$.

Let us now examine the dimension of the set of canonical forms and
deduce the dimension of the generic local equivalence class. First
consider the case where all the individual state spaces have equal
dimension $d$. The number of zero coefficients in the canonical form,
determined by condition 1 of Theorem 1, is $\half nd(d-1)$
(one for each pair $(i,j)$ with $i<j$ and for each position of $j$). The
number of phases removed by condition 2 is $n(d-1)+1$. Hence the
number of real parameters in the canonical form of Theorem 1 is 
\[
 2d^n - nd(d-1) - n(d-1) -1 = 2d^n - n(d^2-1) - 1.
\]

We have not proved that states with different canonical forms are not
locally equivalent; it is conceivable that the number of parameters
could be reduced still further by local transformations. However,
the difference between the number of parameters in the final canonical
form and the dimension of the pure state space must be the dimension of
the generic equivalence class, which is therefore at least $n(d^2-1)+1$.
But this is the dimension of the group of local unitary transformations,
which can be identified with $SU(d)^n\times U(1)$ if we collect together
multiples of the identity in the final $U(1)$. Since the local
equivalence classes are orbits of this group, their dimension cannot be
greater than the dimension of the group. Thus we have
\begin{corollary}
If $n\geq 3$, the generic local equivalence class for a system of $n$ $d$-state particles 
has dimension $n(d^2-1)+1$.
\end{corollary}
Since a generic orbit has the same dimension as the group, the
stabiliser of any state in the orbit has dimension zero. This gives
\begin{corollary} The generic pure state of a system of more than two 
equal-spin particles has a discrete stabiliser under the
action of local unitary transformations.
\end{corollary}

The stabilisers of states of three qubits ($d_1=d_2=d_3=2)$ were studied
in \cite{Hilaryme}. The second corollary generalises Theorem 1 of that
paper. This treatment, however, gives no indication of which exceptional states
have enlarged stabilisers. This question is being investigated in an
alternative approach by one of us (HAC).

For the general situation we must distinguish between the cases $d_n \leq D$
and $d_n > D$ where $D = d_1 \ldots d_{n-1}$. In both cases, the number
of zero coefficients imposed by conditions 1--2
of Theorem 2 is 
\begin{align*}
&\sum_{i=1}^{d_1-1}\sum_{s=1}^n(d_s-i) + 
\sum_{r=1}^{n-2}\sum_{i=d_r}^{d_{r+1}-1}\sum_{s=r+1}^n(d_s-i)\\
=& (d_1-1)\sum_{s=1}^nd_s + \sum_{r=1}^{n-2}(d_{r+1} -
d_r)\sum_{s=r+1}^nd_s
-nS_1 - \sum_{r=1}^{n-2}(n-r)(S_{r+1}-S_r)\\
&\hspace{8cm} \text{ where } S_r=\half d_r(d_r-1)\\
=& -\sum_{s=1}^nd_s + \sum_{r=1}^{n-1}d_r^2 - \sum_{r=1}^{n-1}S_r - S_{n-1}
+ d_n d_{n-1}\\
=&\half\sum_{r=1}^{n}d_r(d_r-1) - \half \delta(\delta +1) 
    \qquad \text{where } \delta =d_n - d_{n-1}.
\end{align*}
If $d_n \leq D$, the number of zero coefficients imposed 
by condition 3 is 
$\half \delta(\delta+1)$, while the number of phases removed by condition 4 is
\[ 
\sum_{r=1}^{n}(d_r-1) + 1.
\]
Hence the number of real parameters in the above canonical form is 
\[
2\prod_{r=1}^n d_r - \(\sum_{r=1}^n (d_r^2 -1) + 1\),
\]
which is the difference between the dimension of the state space and the
dimension of the group $G$ of local transformations. Thus in this case
the dimension of the generic local equivalence class is the same as the
dimension of $G$, as in Corollary 1.

If $d_n > D$, the number of zero coefficients imposed by condition
3 is 
\[ 
\sum_{k=1}^N(\delta - k+1) = N(\delta+1) - \half N(N+1)
\]
giving a total number of zero coefficients determined by conditions
1--3 as
\begin{align*}
&\half\sum_{r=1}^{n}d_r(d_r-1) - \half \delta(\delta + 1) +
\half N(2\delta - N + 1)\\
=&\half\sum_{r=1}^n d_r(d_r - 1) - \half \Delta(\Delta-1) 
\end{align*}
where $\Delta = \delta - N + 1 = d_n - D$. 
In this case there are no non-zero coefficients
$c_{i_1\cdots i_n}$ with $i_n > D$, so the number of phases removed by
condition 4 is reduced by $\Delta$. Hence the total number of parameters
removed, i.e. the dimension of the orbit, is at least
\[ \sum_{r=1}^n (d_r^2 - 1) + 1 - \Delta^2 = \dim G - \Delta^2
\]
where $G = SU(d_1)\times\cdots\times SU(d_n)\times U(1)$.

The fact that there are no  non-zero coefficients
$c_{i_1\cdots i_n}$ with $i_n > D$ means that the state is unaffected by
unitary transformations of the $n$th particle which fix the first $D$
basis vectors. Thus the stability group of the state contains at least 
this $U(\Delta)$ subgroup, and the dimension of the orbit cannot be greater than
dim$G - \Delta^2$. It follows that the dimension of the orbit is exactly
this, and the Lie algebra of the stability group is exactly that of
$U(\Delta)$.

Thus the general versions of Corollaries 1 and 2 are

\begin{corollary}
In the general $n$-party system of Theorem 2, where $n\geq 3$, the generic orbit has
dimension 
\begin{align*}
\sum_{r=1}^n (d_r^2 -1) + 1 \qquad &\text{\emph{if}}\quad d_n \leq D,\\
\sum_{r=1}^n (d_r^2 -1) + 1 - (d_n - D)^2 \quad &\text{\emph{if}}\quad 
d_n > D,
\end{align*}
where $D=d_1\cdots d_{n-1}$.
\end{corollary}
\begin{corollary} 
In the $n$-party system of Theorem 2, a generic point has discrete stabiliser
in the group of local unitary transformations if $d_n \leq D$; otherwise
the stabiliser is locally isomorphic to the unitary group $U(d_n - D)$.
\end{corollary}

\section{Alternative generalisations of the Schmidt decomposition}

The canonical form of the previous section is noteworthy for the
simplicity of the conditions on the coefficients in the expansion of the
state vector in terms of a factorisable orthonormal basis: certain
coefficients are zero, certain others are real. This is likely to make
it most useful in practice. However, it is perhaps less theoretically
appealing than some other generalisations of the Schmidt decomposition
that have been proposed recently. For the sake of completeness, we
review these alternatives here. 

For each constituent of the multipartite system, a basis of its
individual state space is determined by its marginal density matrix:
this is the basis defined by the conventional Schmidt decomposition of
the state vector when the multipartite state space is regarded as a
bipartite tensor product, one factor being the state space of the
constituent being considered, the other being the tensor
product of all the other state spaces. One of us \cite{localinv} and
Brun and Cohen \cite{BrunCohen} have proposed that the tensor products
of these one-particle states provide a natural basis for multipartite
states. The resulting coefficients $c_{i_1\cdots i_n}$ satisfy 
\[ 
\sum_{i_1 \ldots i_{r-1}, i_{r+1} \ldots i_n} 
c_{i_1\cdots i_r\cdots i_n}\overline{c_{i_1\cdots i_{r-1}j_ri_{r+1}\cdots i_n}} = 0 
\quad\text{if}\quad i_r\ne j_r
\]
for each $r$.

Spekkens and Sipe \cite{SpekkensSipe} have suggested that a
canonical state in each equivalence class could be taken to be that
which minimises the Ingarden-Urbanik entropy
\[
S_{\text{IU}} = - \sum_{i_1\cdots i_n} |c_{i_1\cdots i_n}|^2\log
|c_{i_1\cdots i_n}|^2.
\]
They justify this as a generalisation of the Schmidt decomposition by
showing that the IU entropy is minimised by the Schmidt normal form for
bipartite states. For more than two constituent parts, however, little
is known about these minima. 

To show that all three of these canonical forms are distinct, consider the
tripartite state
\[ 
|\Psi\> = \frac{1}{2\sqrt{3}}\(
3|000\> + |011\> + \sqrt{2}|111\>\)
\]
where each constituent system is a qubit (a two-state system) and, as
usual, we label the basis states by the digits 0 and 1 and abbreviate
the product basis states as 
$|abc\>=|\psi^{(1)}_a\>|\psi^{(2)}_b\>|\psi^{(3)}_c\>$.
This state is presented in the canonical form of section 2; not only
does it satisfy the conditions of Theorem 1, but it can be shown (see
Appendix) that it is obtained by the procedure of that theorem, i.e. the
coefficient of $|000\>$ is maximal among states locally equivalent to
$|\Psi\>$. (We note that there is a locally equivalent state with the
coefficient of $|000\>$ equal to $\frac{1}{2}$; this satisfies
conditions 1 and 2 of Theorem 1, but not condition 3.) 
By means of a transformation of the first qubit with matrix 
$\frac{1}{\sqrt{6}}\(\begin{smallmatrix}\sqrt{2}+1&1-\sqrt{2}
\\\sqrt{2}-1&\sqrt{2}+1\end{smallmatrix}\)$), the state $|\Psi\>$ is
locally equivalent to
\[
|\Phi\> =
\frac{1}{2\sqrt{2}}\((\sqrt{2}+1)|000\>-(\sqrt{2}-1)|100\>
+ |011\> + |111\>\)
\]
which is in the canonical form of \cite{localinv} and \cite{BrunCohen}.
Neither of these states minimises the IU entropy within their
local equivalence class: under the infinitesimal local transformation in
which the basis states of the first qubit transform by $|0\>\mapsto
|0\> + \epsilon |1\>, |1\>\mapsto |1\> -\epsilon |0\>$ and the states 
of the second and third qubits
are kept fixed, the entropies of $|\Psi\>$ and $|\Phi\>$ change by 
\[ 
\delta S_{\text{IU}}(\Psi) = -\tfrac{1}{3\sqrt{2}}\epsilon\log 2, \qquad
\delta S_{\text{IU}}(\Phi) = -\epsilon\log(\sqrt{2}+1).
\]
By minimising the IU entropy numerically we find in general
that a few of the coefficients $c_{ijk}$ become much
smaller than the others without vanishing exactly.

\section*{Acknowledgement}

We are grateful to Bob Gingrich for pointing out an error in an earlier
version of this paper.

\newpage
\section*{Appendix}

We will show that for a state 
\[
|\Psi\> = a|000\> + b|011\> + c|111\>
\]
with $|a| > \frac{1}{\sqrt{2}}$, $a$ is the maximal value of the
coefficient of $|000\>$ among states locally equivalent to $|\Psi\>$.

\begin{proof}
The equations \eqref{E-L} for a stationary value of 
$\left|\<\Psi|\(|\phi^{(1)}\>|\phi^{(2)}\>|\phi^{(3)}\>\)\right|^2$ become
\begin{align}
\label{1} \overline{a}v_0 w_0 +\overline{b}v_1 w_1 & = \lambda \overline{u_0}\\
\label{2} \overline{c}v_1 w_1 &= \lambda \overline{u_1}\\
\label{3} \overline{a}u_0 w_0 &= \lambda \overline{v_0}\\
\label{4} \overline{b}u_0 w_1 + \overline{c}u_1 w_1 &= \lambda\overline{v_1}\\
\label{5} \overline{a}u_0 v_0 &= \lambda\overline{w_0}\\
\label{6} \overline{b}u_0 v_1 + \overline{c}u_1 v_1 &= \lambda \overline{w_1}
\end{align}
where $|\phi^{(1)}\> = u_0|0\> + u_1|1\>$,
$|\phi^{(2)}\> = v_0|0\> + v_1|1\>$ and
$|\phi^{(3)}\> = w_0|0\> + w_1|1\>$.
Clearly there is a solution $\lambda = \overline{a}, u = v = w = (1,0)$.
We have to
show that any other solution has $|\lambda|^2 < |a|^2$.  Using \eqref{5}
and \eqref{6} to eliminate $w_0$ and $w_1$, eqs. \eqref{1} and \eqref{2}
become
\begin{align*}
\(|a|^2|v_0|^2 + |b|^2|v_1|^2 - |\lambda|^2\)u_0 +
b\overline{c}|v_1|^2u_1 &= 0,\\
\overline{b}c|v_1|^2 u_0 + \(|c|^2|v_1|^2 - |\lambda|^2\)u_1 &= 0.
\end{align*}
For $(u_o, u_1) \ne (0,0)$, it follows that 
\[
F(|\lambda|^2) = |\lambda|^4 - |\lambda|^2\( |a|^2 + (1 - 2|a|^2)|v_1|^2\) +
|a|^2|c|^2|v_0|^2|v_1|^2 = 0
\]
since $|a|^2 + |b|^2 +|c|^2 = |v_0|^2 +|v_1|^2 = 1$. Now 
\[
F(|a|^2) =|a|^2|v_1|^2\(2|a|^2 -1 + |c|^2|v_0|^2\).
\]
which is positive if $|a|>\frac{1}{\sqrt{2}}$ (unless $v_1=0$), and the gradient of the
quadratic $F$ at $|\lambda|^2 = |a|^2$ is also positive. It follows that
the zeros of $F$, and therefore any stationary
values of $\left|\<\Psi|\(|\phi^{(1)}\>|\phi^{(2)}\>|\phi^{(3)}\>\)\right|^2$
other than $|a|^2$, are less than $|a|^2$.
\end{proof}

\end{document}